\documentclass[conference]{IEEEtran}
\IEEEoverridecommandlockouts
\usepackage{cite}
\usepackage{amsmath,amssymb,amsfonts}
\usepackage{algorithmic}
\usepackage{graphicx}
\usepackage{textcomp}
\usepackage{xcolor}
\usepackage{siunitx}
\usepackage[capitalise]{cleveref}
\usepackage{lipsum}
\usepackage{nicefrac}
\usepackage{tikz}
\usepackage[nolist]{acronym}
\usepackage[keeplastbox]{flushend}
\def\BibTeX{{\rm B\kern-.05em{\sc i\kern-.025em b}\kern-.08em
    T\kern-.1667em\lower.7ex\hbox{E}\kern-.125emX}}

\newcommand{\ntones}{K}
\newcommand{\tonespacing}{\Delta f}
\newcommand{\txtonespacing}{\Delta f_{\text{Ant}}} 
\newcommand{\nsnap}{N_{\text{snap}}} 
\newcommand{\txidx}{i} 

\begin{document}
\title{Sparsity in the Delay-Doppler Domain for %
Measured 60 GHz Vehicle-to-Infrastructure Communication Channels%
\thanks{%
The support for this research by Wien Leuchtet (Magistratsabteilung 33) is gratefully acknowledged. %
This work was supported in part by the Austrian Federal Ministry for Digital and Economic Affairs, %
in part by the National Foundation for Research, Technology and Development, %
in part by the Czech Science Foundation under Project 17-27068S, %
and in part by the National Sustainability Program under Grant LO1401. %
The work of S.~Sangodoyin and A.~F.~Molisch was supported in part by NSF under Grant CNS-1457340, %
and in part by NIST under Grant 70NANB17H157. %
This work was carried out in the framework of COST Action CA15104 IRACON.%
}
}

\author{%
	\IEEEauthorblockN{
	Herbert~Groll\IEEEauthorrefmark{1}, %
	Erich~Z\"{o}chmann\IEEEauthorrefmark{1,}\IEEEauthorrefmark{2,}\IEEEauthorrefmark{3}, %
	Stefan~Pratschner\IEEEauthorrefmark{1,}\IEEEauthorrefmark{2}, %
	Martin~Lerch\IEEEauthorrefmark{1}, \\%
	Daniel~Sch\"{u}tzenh\"{o}fer\IEEEauthorrefmark{1,}\IEEEauthorrefmark{2}, %
	Markus~Hofer\IEEEauthorrefmark{4}, %
	Jiri~Blumenstein\IEEEauthorrefmark{3}, %
	Seun~Sangodoyin\IEEEauthorrefmark{5}, \\%
	Thomas~Zemen\IEEEauthorrefmark{4}, %
	Ale\v{s}~Proke\v{s}\IEEEauthorrefmark{3}, %
	Andreas~F.~Molisch\IEEEauthorrefmark{5}, and %
	Sebastian~Caban\IEEEauthorrefmark{1,}\IEEEauthorrefmark{2} %
}\\
\IEEEauthorblockA{\IEEEauthorrefmark{1} Institute of Telecommunications, TU Wien, Austria}
\IEEEauthorblockA{\IEEEauthorrefmark{2} Christian Doppler Laboratory for Dependable Wireless Connectivity for the Society in Motion}
\IEEEauthorblockA{\IEEEauthorrefmark{3} Department of Radio Electronics, TU Brno, Czech Republic}
\IEEEauthorblockA{\IEEEauthorrefmark{4} Center for Digital Safety \& Security, AIT Austrian Institute of Technology, Austria}
\IEEEauthorblockA{\IEEEauthorrefmark{5} Wireless Devices and Systems Group, University of Southern California, USA}
}

\maketitle

\begin{abstract}
	We report results from millimeter wave vehicle-to-infrastructure (V2I) channel measurements conducted on Sept.~25,~2018 in an urban street environment, down-town Vienna, Austria. %
	Measurements of a frequency-division multiplexed multiple-input single-output
	channel have been acquired with a time-domain channel sounder at 60 GHz
	with a bandwidth of 100 MHz and a frequency resolution of 5 MHz. %
	Two horn antennas were used on a moving transmitter vehicle: one horn emitted a beam towards the horizon and the second horn emitted an elevated beam at 15-degrees up-tilt. %
	This configuration was chosen to assess the impact of beam elevation on V2I communication channel characteristics: propagation loss and sparsity of the local scattering function in the delay-Doppler domain. %
	The measurement results within urban speed limits show high sparsity in the delay-Doppler domain. %
\end{abstract}
\begin{IEEEkeywords}
OTFS, delay-Doppler domain, sparsity, millimeter wave, vehicular communications, channel estimation
\end{IEEEkeywords}
\acresetall
\section{Introduction}
Vehicular communications was intensely investigated in the last decade~\cite{cheng2007mobile,renaudin2008wideband,paier2009characterization,mecklenbrauker2011vehicular}.
MmWave have been a research topic in vehicular communications for already several decades\cite{meinel1983millimetre,kato2001propagation}.
The path loss for train-to-infrastructure scenarios in the \ac{mmWave} frequency band is measured in \cite{meinel1983millimetre}, performance for \ac{V2V} communication is investigated in \cite{kato2001propagation}.

The effect of an overtaking vehicle on a \ac{V2V} channel at \SI{60}{\giga\hertz} of a communicating car platoon has been investigated in~\cite{zochmann2018measured,blumenstein2018measured}. 
A more detailed statistical analysis of delay and Doppler spread is provided in~\cite{zochmann2018statistical,zochmann2019position}. 
\ac{V2V} channel measurements with antennas placed in the bumpers of cars at \SI{38}{\giga\hertz} and \SI{60}{\giga\hertz},
using a channel sounder with \SI{1}{\giga\hertz} bandwidth, have been
conducted in \cite{sanchez2017millimeter}.
Further \ac{mmWave} \ac{V2V} measurements with approaching cars are shown in~\cite{prokes2018timedomain}.
Doppler spectra of \ac{V2I} measurements at \SI{28}{\giga\hertz} in an expressway environment are shown in\cite{park2018v2i}.
Doppler spectra of vibrations appearing while the vehicle
is in operation are shown in \cite{prokes2016time} and \cite{blumenstein2017time}.

The \ac{mmWave} wireless channel has been observed to have a sparse multipath structure in multiple domains~\cite{sayeed2007maximizing,gustafson2014mm,Zochmann2019_Eurasip}.
For channel estimation using sparse reconstruction, several techniques exist~\cite{venugopal2017channel}.
Sparse channel fits via the least absolute shrinkage and selection operator (LASSO) applied in the delay domain are provided in~\cite{blazek2017sparse,blazek2018approximating,blazek2018model,blazek2018millimeter}.
In~\cite{blazek2018model} and~\cite{blazek2018millimeter}, Akaike's information criterion was used to select the appropriate sparsity level, where approximately 4 to 6 clusters in the delay domain have been found.
Sparse signal reconstruction using the \ac{SBL} framework with low-complexity is used in~\cite{prasad2014joint,gerstoft2016multisnapshot}.
A fixed-point implementation of \ac{SBL}~\cite{gerstoft2016multisnapshot} is shown in~\cite{groll2018sparse}.

How to exploit the delay-Doppler sparsity of wireless channels is shown in~\cite{taubock2008compressed}.
The \ac{OTFS} modulation~\cite{hadani2017orthogonal} performs especially well for wireless channels with sparsity in the delay-Doppler domain, such as in \ac{mmWave} communication systems\cite{hadani2017mmwave}.
\section*{Our contribution}
In this contribution, we show that the sparsity in the delay-Doppler domain holds true for measured \SI{60}{\giga\hertz} \ac{V2I} communication channels in a street crossing scenario in an urban environment.
We estimate the channel and find a representation in the delay-Doppler domain with a multitaper estimator.
Furthermore, a sparse estimation with sub-Nyquist sampling in the delay domain is carried out using \ac{SBL}.

\section{Measured Scenario}
\begin{figure}[htb]
	\begin{center}%
	\begin{tikzpicture}
	\node (img1) {%
		\includegraphics{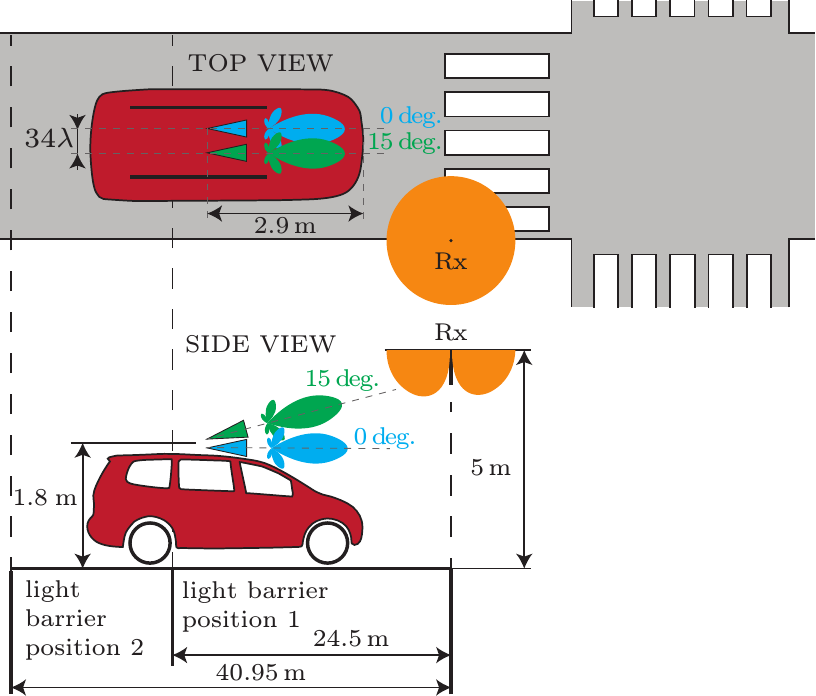}%
	};
	\node[anchor=south east, yshift=0.5cm] (img2) at (img1.south east) {\includegraphics[width=2.8cm]{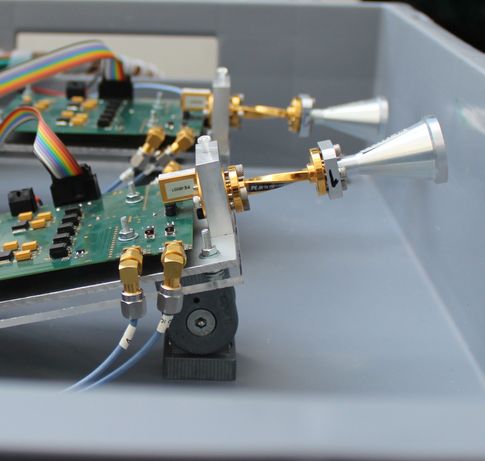}};
	\node[anchor=north west, yshift=+0.1cm] at (img1.north west) {(a)};
	\node[anchor=south east, yshift=+0.1cm] at (img2.south east) {(b)};
	\end{tikzpicture}
\end{center}%
\vspace{-0.5cm}
\caption{(a)~Measurement scenario with transmitter car, and receiver at crossroads. (b)~Transmitter modules with different antenna elevation.}%
\label{fig:scenario}%
\end{figure}

The measurement scenario and placement of the \acp{TX} and the \ac{RX} is shown in \cref{fig:scenario}{a}.
Our \acp{TX} are mounted on the roof rack of a car.
The \ac{RX} is positioned at the crossroads of a street canyon, at \SI{5}{\meter} height.
We are observing \acp{MPC} of the wireless channel during the approach and passage of the moving transmitter vehicle at the crossroads.
All \acp{MPC} are Doppler shifted proportional to their respective radial velocity.

For our setup, the \SI{60}{\giga\hertz} free space path loss calculates to \SI{100}{\decibel} at the maximum distance of interest.
Both \acp{TX} use horn antennas with a gain of \SI{20}{\decibel i}, which point towards driving direction.
One horn (TX~\ang{0}) emits a beam towards the horizon and the second horn (TX~\ang{15}) emits a beam at \ang{15} up-tilt as shown in \cref{fig:scenario}{b}.
At the receiver side, an omni-directional $\nicefrac{\lambda}{4}$ monopole antenna is used.
The receive antenna gain is approximately \SI{-4}{\decibel i} including cable losses.
%
\section{RF Setup}
\begin{figure}[htb]
	\begin{center}%
	\includegraphics{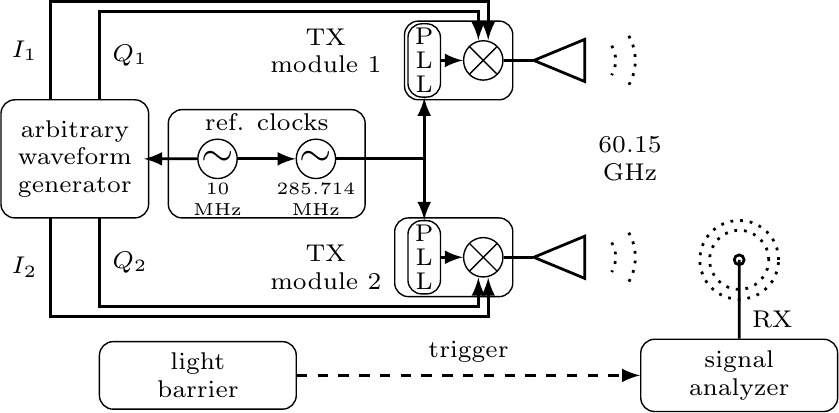}%
	\end{center}%
	\vspace{-0.3cm}
	\caption{Measurement setup.}
	\label{fig:hwsetup}
\end{figure}

The hardware setup for the measurement is shown in \cref{fig:hwsetup}.
Both \acp{TX} consist of external in-phase quadrature (I/Q) mixer modules, an \ac{AWG}, and a frequency reference.
The \ac{AWG} continuously repeats a baseband sounding sequence as I/Q signal for each external I/Q-mixer module.
Each I/Q-mixer up-converts the input signal to passband at \SI{60.15}{\giga\hertz} center frequency with a \acl{LO}, synthesized by a \ac{PLL}.
The \acl{RF} outputs of the I/Q-mixers emit \acp{mmWave} through conical horn antennas.
The frequency reference feeds each \ac{PLL} with a \SI{285.714}{\mega\hertz} reference and the \ac{AWG} with a \SI{10}{\mega\hertz} reference.
Our receiver is a \ac{SA} connected to the omni-directional antenna and a light barrier.
We measure the \ac{CFO} between \ac{AWG} and SA at standstill before each measurement.
The oscillator drift during one measurement was found to be less than \SI{1}{\hertz}.
A measurement is started as soon as the moving transmitter vehicle passes the light barrier, which triggers the recording at the \ac{RX}.
We directly access the I/Q samples from the SA with a sampling rate of $\SI{125}{\mega Samples \per \second}$.
\section{Sounding Sequence Design}
We extract an estimate of the channel transfer function with a multitone waveform as excitation signal for channel sounding similar to the approach in \cite{sangodoyin2012real}.
A \acl{ZC} sequence~\cite{chu1972polyphase} is used for its \acl{CAZAC} property to keep the crest factor low and to phase-align both \acp{TX} during calibration.
For our scenario, we consider a maximum excess delay $\tau_{\text{max}}$ to the assumed propagation delay between the \ac{LOS} component and \acp{MPC} due to single reflections.
Paths due to multiple reflections and paths due to single reflections, due to back- and side-lobes at the \acp{TX}, are below the receiver sensitivity.
Possible single reflections are limited by the surrounding street canyon.
The width of the street canyon is \SI{20}{\meter}.
Therefore, at a maximum distance of \SI{44}{\meter} between \acp{TX} and \ac{RX}, we account for path differences between \ac{LOS} and \acp{MPC}, which will not be larger than \SI{16}{\meter} in length or \SI{54}{\nano\second} in time.
Reflections at the crossing and beyond the crossing might lead to delays larger than $\SI{100}{\nano\second}$.
The tone spacing $\tonespacing$ of our multitone waveform has to fulfill the sampling theorem.
Therefore, $\tonespacing \leq{1\slash (2\tau_{\text{max}})}$ needs to hold true for the maximum excess delay $\tau_{\text{max}}$ of the channel.
We use a measurement bandwidth $B=\SI{100}{\mega\hertz}$ and $\ntones=\num{21}$ tones to get a tone spacing of $\tonespacing = \SI{4.76}{\mega \hertz}$,
which leads to a maximum alias free delay  of $\tau_{\text{max}} = \SI{105}{\nano \second}$.

To measure the channel simultaneously for our two \acp{TX}, we use a frequency division multiplexing (FDM) approach as described in \cite{konishi2011channel}.
We derive the multitone waveform for the second \ac{TX} from the first \ac{TX} with a frequency offset $\txtonespacing=\SI{1.19}{\mega\hertz}$ for every single tone.
The superposition of both waveforms leads to a multitone-comb, so that \acp{TX} are orthogonally interleaved in frequency.
The resulting multitone sounding sequence for each \ac{TX} has a period of $T=\SI{840}{\nano\second}$, leaving possible additional tones unoccupied.

The snapshot time $T_{\text{snap}}$ for the channel has to fulfill the sampling theorem for a maximum Doppler frequency $\nu_{\text{max}}$, so that $T_{\text{snap}} \leq {1 \slash (2 \nu_{\text{max}})}$.
For a maximum assumed speed of approaching cars $v_{\text{max}} = \SI{14}{\meter \per \second}$, %
$\nu_{\text{max}} = \SI{2800}{\hertz}$ and thus $T_{\text{snap}} \leq {1 \slash (2 \nu_{\text{max}})} = \SI{178.6}{\micro \second}$.

The final \ac{SNR} is enhanced to combat thermal noise by coherently averaging over $N$ channel measurements up-to the maximum snapshot time.
By using $N=212$, a processing gain of $\SI{23.26}{\deci\bel}$ is achieved and $T_{\text{snap}}=\SI{178.07}{\micro\second}$ is low enough for $\nu_{\text{max}}$.

Our \ac{RX} is able to record up to \SI{450}{\mega Samples} at one shot, which influences the choice of measurement bandwidth. For a sampling rate of \SI{125}{\mega Samples \per \second}, the recording time is $T_{\text{rec}}=\SI{3.6}{\second}$ which translates to a travel distance of \SI{50.4}{\meter} at $v_{\text{max}}$.
An example of the measurement run is shown in \cref{sec:results}.
The parameters of the channel sounder are summarized in \cref{tab:parameters}.
\begin{table}[htbp]
	\vspace{-0.4cm}
	\caption{Channel sounder parameters}
	\begin{center}
		\begin{tabular}{rl}
			\hline
			\textbf{Parameter} & \textbf{Value} \\
			\hline
			center frequency & $f_c=\SI{60.15}{\giga\hertz}$ \\
			tone spacing & $\tonespacing = \SI{4.76}{\mega \hertz}$  \\
			number of tones  & $\ntones = \num{21}$ \\
			number of TX antennas & $2$ \\
			tone spacing between antennas & $\txtonespacing=\SI{1.19}{\mega\hertz}$ \\
			bandwidth & $B=\SI{100}{\mega\hertz}$ \\
			maximum alias free delay & $\tau_{\text{max}} = \SI{105}{\nano \second}$ \\
			transmit antennas & \SI{20}{\decibel i} conical horn \\
			receive antenna & $\nicefrac{\lambda}{4}$ monopole \\
			snapshot time & $T_{\text{snap}}=\SI{178.07}{\micro\second}$ \\
			maximum car speed & $v_{\text{max}} = \SI{14}{\meter \per \second}$ \\
			recording time & $T_{\text{rec}}=\SI{3.6}{\second}$ \\
			\hline
		\end{tabular}
		\label{tab:parameters}
	\end{center}
\end{table}
\newcommand{\fidx}{f}
\newcommand{\tidx}{t}
\newcommand{\allfidx}{q} 
\newcommand{\nallf}{Q} 
\newcommand{\evaltime}{M} 
\section{Measurement Evaluation}
Frequency and time synchronization is performed utilizing the \ac{LOS} component.
\paragraph*{Frequency synchronization} 
The \ac{CFO} between the oscillator of \ac{RX} and \ac{TX} is measured at standstill with a correlator before each measurement.
\paragraph*{Coherent averaging}
I/Q samples of the actual measurement with the car in motion are recorded.
A frequency offset of the \ac{LOS} due to Doppler plus \ac{CFO} is estimated using correlators for each \ac{TX}.
We calculate the average received sounding sequence by applying a frequency offset correction and coherent averaging.
The frequency correction, without the \ac{CFO}, is reversed afterwards.
For averaging, we use a rectangular window without overlap.
\paragraph*{Time synchronization through LOS alignment}
We estimate the propagation delay of the \ac{LOS} component from the known distance between \acp{TX} at the light barrier position to the \ac{RX}.
Therefore, we align the delay of the \ac{LOS} component accordingly. 

We calculate estimates of the time-variant transfer function $H_\txidx [ \tidx, \fidx_\txidx ] $ similar to \cite{zochmann2018measured}, where $\tidx$ is the time index $\tidx=\{\tidx_0, \dots, \tidx_{\nsnap-1}\}$ of a snapshot and $\fidx_\txidx$ are the frequency indices $\fidx_\txidx=\{\fidx_{\txidx,0}, \dots, \fidx_{\txidx,K-1}\}$ selected for the given \ac{TX}~$i$ due to the interleaved multitone waveform.
For a total recording time $T_{\text{rec}}=\SI{3.6}{\second}$, the number of snapshots is $\nsnap=\num{20275}$. 
As common for example in \ac{OTFS} modulation~\cite{hadani2017orthogonal}, we characterize the channel in delay-Doppler domain using a \ac{SFFT} on $H_\txidx [\tidx, \fidx_\txidx]$.
We use a multitaper based estimator within an evaluation region of $\evaltime$ samples in time and $\ntones$ samples in frequency.
The employed tapers are \acl{DPSS}\cite{slepian1978prolate} to allow averaging of multiple independent spectral estimates of the same measurement with confinement in time and frequency.
Furthermore, this type of estimator leads to delay and Doppler confined estimates.
This approach is known as \ac{LSF}\cite{bernado2014delay,bernado2015time,thomson1982spectrum}.
We use three tapers in time domain and three tapers in frequency domain.

From the \ac{LSF}, we derive the \ac{DSD} as explained in\cite{bernado2014delay}.
For evaluation of the \ac{LSF}, we set the number of snapshots to $\evaltime = 360$ and number of frequencies to $\ntones = 21$ which corresponds to an evaluation time of $T_{\text{eval}}=\SI{64.3}{\milli \second}$ and bandwidth of $B_{\text{eval}}=\SI{100}{\mega \hertz}$.
In \cref{fig:dsd} and \cref{fig:delaydoppler} we show the \ac{DSD} and the \ac{LSF} over the recording time.

\paragraph*{Formalism for a sparse representation of the channel}
\newcommand{\timeidx}{l}
\newcommand{\freqidx}{k}
\newcommand{\delayidx}{n}
\newcommand{\doppleridx}{m}
\newcommand{\numfreqdelay}{K}
\newcommand{\numtimedoppler}{M}
We use the \ac{SFFT} to transform consecutive time-variant frequency responses $H_i[\freqidx, \timeidx]$ into a Doppler-variant impulse response $S_i[\delayidx, \doppleridx]$ and vice versa with the \ac{ISFFT}.
We write $H_i[\freqidx, \timeidx]$ and $S_i[\delayidx, \doppleridx]$ as matrix $\mathbf{H}\in\mathbb{C}^{\numfreqdelay \times \numtimedoppler}$ and $\mathbf{S}\in\mathbb{C}^{\numfreqdelay \times \numtimedoppler}$ respectively for each $i$.
$\numfreqdelay$ is the number of frequency indices $\freqidx$ and delay indices $\delayidx$, and $\numtimedoppler$ is the number of time indices $\timeidx$ and Doppler indices $\doppleridx$.
Splitting up the \ac{ISFFT} into a \ac{DFT} and an inverse \ac{DFT} as in \cite{rezazadehreyhani2018analysis} leads to 
\begin{align}\label{eq:ISFFT}
\mathbf{H}=\mathbf{F}_\numfreqdelay \mathbf{S} \mathbf{F}_\numtimedoppler^H ~,
\end{align}
where the index of the \ac{DFT} matrix $\mathbf{F}$ denotes the size of the \ac{DFT}.
Using the \textit{Kronecker} product $\otimes$, we find the vectorization $\text{vec}(\cdot)$ of $\mathbf{H}$ as  
\begin{align}
\mathbf{h}=\left(\mathbf{F}_\numtimedoppler^H \otimes \mathbf{F}_\numfreqdelay\right)\mathbf{s} ~,
\end{align}
with $\mathbf{h}=\text{vec}\left(\mathbf{H}\right)$ and $\mathbf{s}=\text{vec}\left(\mathbf{S}\right)$.
With the transfer matrix $\mathbf{A}=\left(\mathbf{F}_\numtimedoppler^H \otimes \mathbf{F}_\numfreqdelay\right)$ and additive noise $\mathbf{w}$, $\mathbf{h}$ is expressed as
\begin{align}
\mathbf{h} = \mathbf{A}\mathbf{s} + \mathbf{w} ~,
\end{align}
where we want to find a vector $\mathbf{\hat{s}}$ which is a sparse representation of the measured channel.
In this contribution a sparse fit is obtained with two different methods.
\paragraph*{Largest peaks of LSF}
We find a sparse fit through trivial two-dimensional peak search of the \ac{LSF} and selection of the ten largest peaks.
\paragraph*{Sparse fit using SBL}
Moreover, we find a sparse representation of the channel $\mathbf{h}$ with the \ac{SBL} algorithm in~\cite{gerstoft2016multisnapshot} as a solver.
Based on $\mathbf{F}_\numfreqdelay$, we define a \ac{DFT} matrix $\mathbf{\tilde{F}}_\numfreqdelay \in \mathbb{C}^{\numfreqdelay \times 4\numfreqdelay}$ for super-resolution in the delay domain by a factor of $4$.
This enhances the resolution of the estimation, which is given by the channel sounder's sampling in frequency.
The modified transfer matrix for the solver is then
$\mathbf{\tilde{A}}=\mathbf{F}_\numtimedoppler^H \otimes \mathbf{\tilde{F}}_\numfreqdelay \in \mathbb{C}^{\numfreqdelay\numtimedoppler \times 4\numfreqdelay\numtimedoppler}$, which leads to the model
\begin{align}
\mathbf{h} = \mathbf{\tilde{A}}\mathbf{\tilde{s}} + \mathbf{w} ~,
\end{align}
with $\mathbf{\tilde{s}} \in \mathbb{C}^{4\numfreqdelay\numtimedoppler}$, the resolution enhanced version of $\mathbf{s}$.
Instead of $\mathbf{\tilde{s}}$, \ac{SBL} iteratively estimates the channel powers through the variances $\boldsymbol{\gamma}=[\gamma_1,\dots,\gamma_{4 \numfreqdelay \numtimedoppler}]^T$.
Furthermore, each iteration in \ac{SBL} estimates the noise variance based on active set selection using the $P$ largest peaks in $\boldsymbol{\gamma}$.
We modify the one-dimensional selection in~\cite{gerstoft2016multisnapshot} for a two-dimensional peak selection in $\text{vec}^{-1}\left(\boldsymbol{\gamma}\right) \in \mathbb{R}^{4 \numfreqdelay \times \numtimedoppler}$ according to our delay-Doppler plane.
For the \ac{SBL} algorithm, we use the SBL1 variant in~\cite{gerstoft2016multisnapshot} and $P=10$.
Initial channel variances are $\boldsymbol{\gamma}=[1,\dots,1]^T$ and initial noise variance is $0.1$.
Ten iterations are done.
\section{Results}\label{sec:results}
\begin{figure}[htb]
	\begin{center}%
		\includegraphics[width=0.485\textwidth]{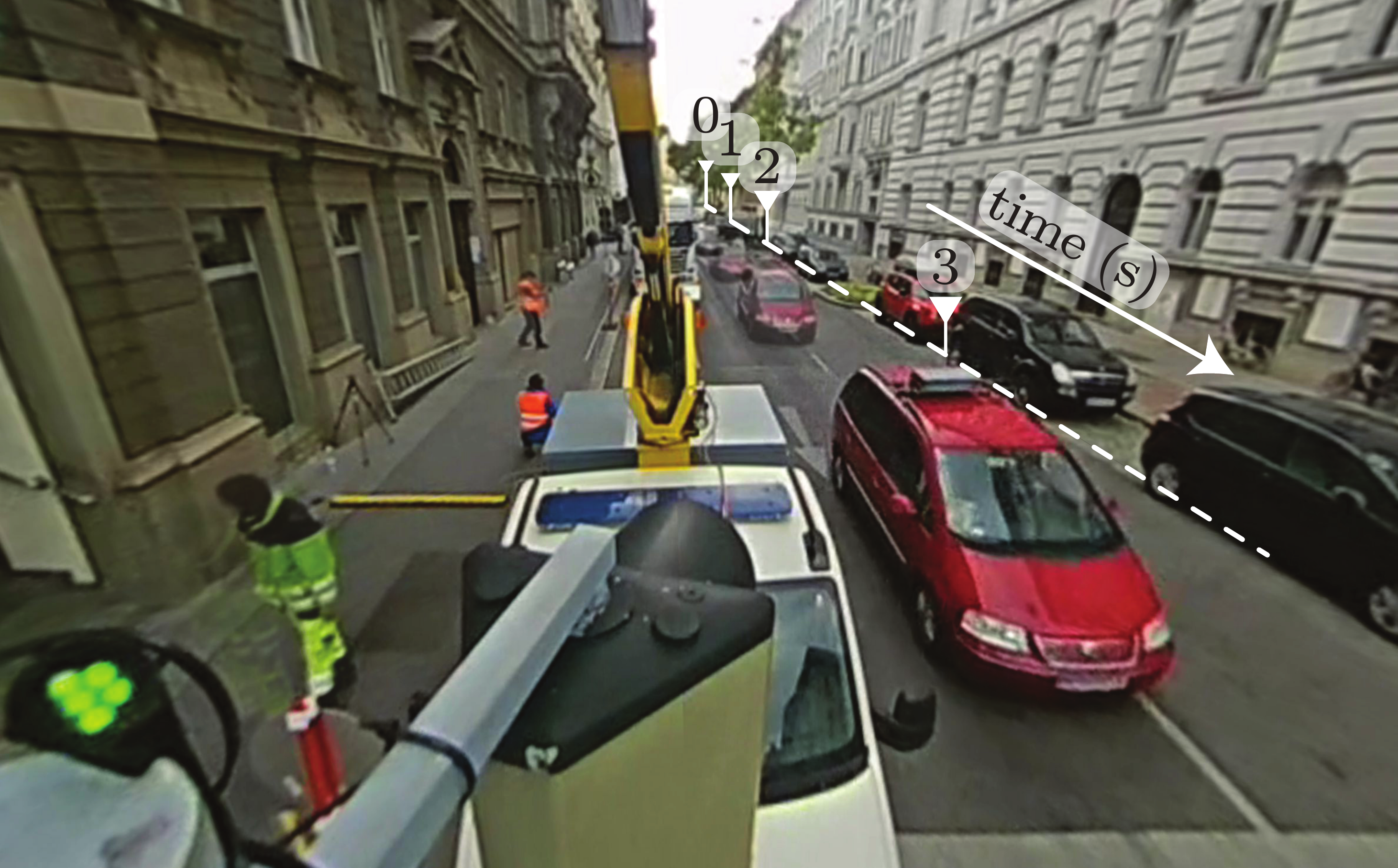}
	\end{center}%
	\vspace{-0.3cm}
	\caption{Our transmitter car passing through the measurement scenario. %
		The car triggers the recording by passing the light barrier. %
		Within the first second of the recording, additional \acfp{MPC} exist due to a parked truck, which contribute to fading.
		Those \acp{MPC} no longer exist after the car passes the truck at $\SI{2}{\second}$.
		The car passes the receiver near $\SI{3}{\second}$.
	}
	\label{fig:measurement_example}
\end{figure}
\begin{figure}[htb]
	\vspace{-0.3cm}
	\begin{center}%
		\includegraphics[width=0.485\textwidth]{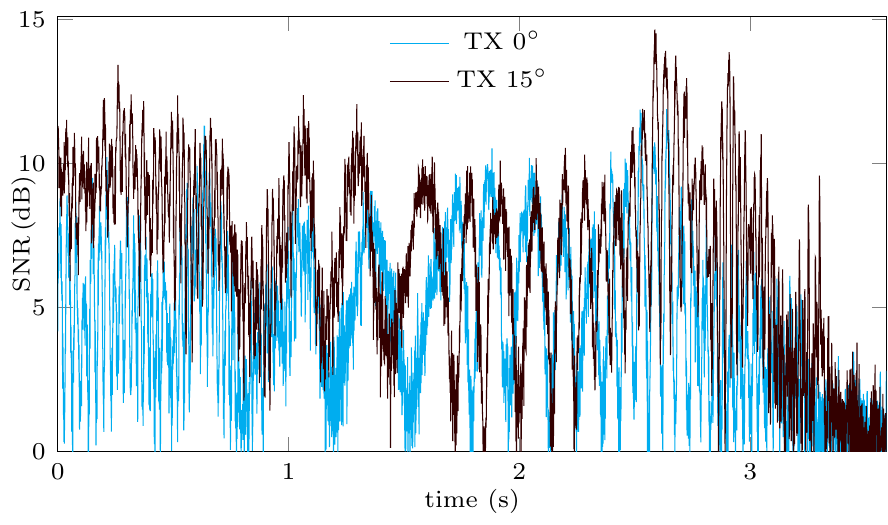}
	\end{center}%
	\vspace{-0.5cm}
	\caption{The receiver SNR per transmitter for the measurement example. %
		Additional \acp{MPC} in the first third of the recording cause stronger fading for TX~\ang{0} and more SNR for TX~\ang{15}. %
		Both \acp{TX} show similar SNR in the second third because fading is dominated by two signal paths.
		In the last third, the elevated TX~\ang{15} shows a short soar in SNR compared to TX~\ang{0} due to the spatial filtering.
		The SNR drops quickly for both \acp{TX} as the car approaches the area below the receiver. %
	}
	\label{fig:rx_power}
\end{figure}
\begin{figure}[htb]
	\begin{center}%
		\includegraphics{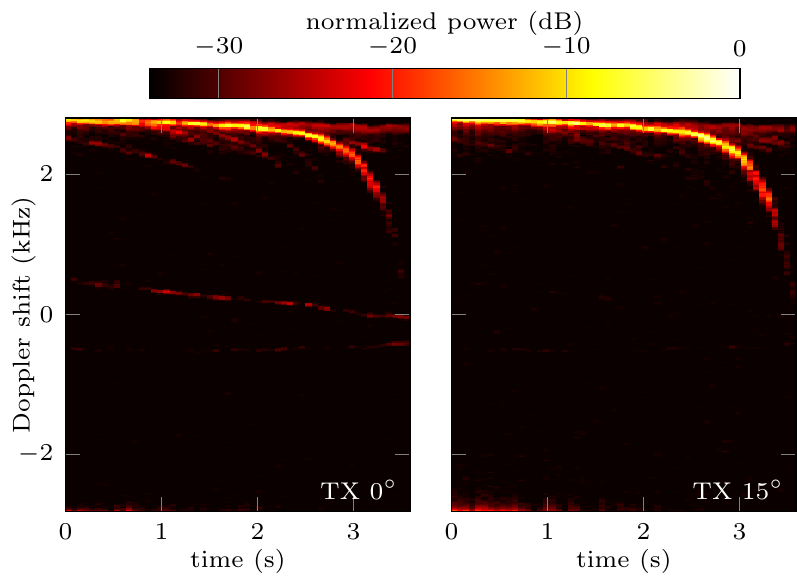}%
	\end{center}%
	\vspace{-0.7cm}
	\caption{
		Doppler power spectral density of the measurement example.
		At $\SI{0}{\second}$, the \ac{LOS} has a Doppler-shift of more than \SI{2.5}{\kilo\hertz}.
		The \ac{LOS} dominates the wireless channel in the Doppler domain equally for both \acp{TX}.
		The TX~\ang{15} shows fewer \acp{MPC} due to spatial filtering of the street level.
		Doppler-shift of the \ac{LOS} moves towards negative frequencies due to change in radial velocity as the car passes the receiver.
	}%
	\label{fig:dsd}%
	\vspace{-0.2cm}%
\end{figure}
As an example, we show a measurement in \cref{fig:measurement_example} with the transmitter car passing through the whole measurement scenario at constant speed.
The recording is triggered when the car passes the light barrier \SI{41}{\meter} before the \ac{RX} position.
The car continues towards the crossroads without stopping at the green traffic light and then passes the \ac{RX}.
The \ac{SNR} at the \ac{RX} is shown in \cref{fig:rx_power}.
Fading is different for both \ac{TX} antennas.
Fading through \acp{MPC} in the first third of the recording is due to a parked truck.
In the second third, fading is equal for both \ac{TX} antennas because only two dominant signal paths exist.
As the car passes the \ac{RX}, \ac{SNR} decreases first for the horizontal beam and later for the elevated beam due to spatial filtering.
The Doppler power spectral density is shown in \cref{fig:dsd} with a dominant \ac{LOS} component.
The elevated beam shows fewer \acp{MPC} due to spatial filtering of the street level.
For both antenna elevations, a \ac{LOS} component with Doppler shift of more than \SI{2.5}{\kilo\hertz} is visible at $\SI{0}{\second}$.
As the car approaches the \ac{RX}, Doppler shift of the \ac{LOS} component decreases towards negative frequencies.

\begin{figure*}
	\begin{center}%
		\includegraphics[scale=1.3]{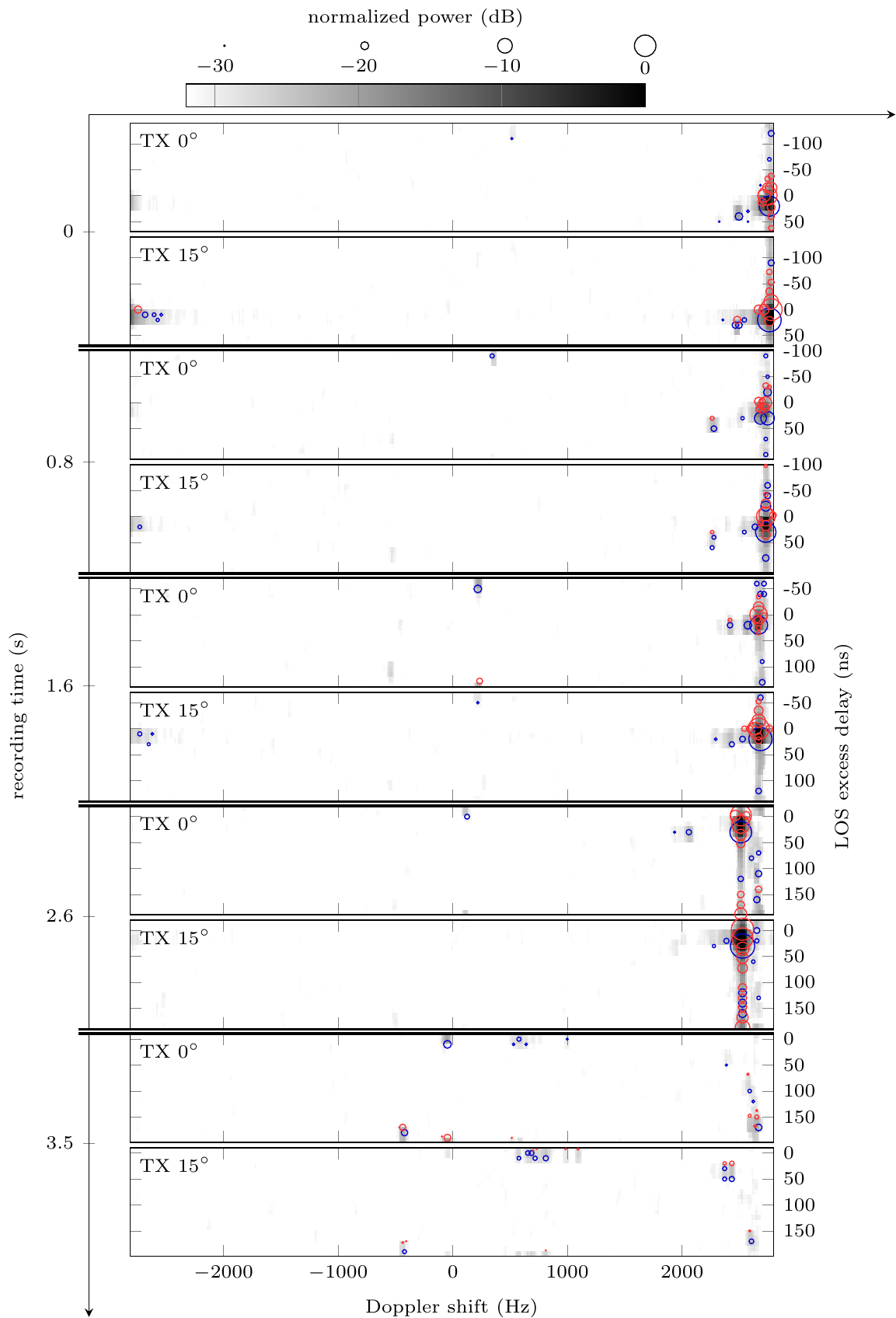}%
	\end{center}%
	\caption{%
		Spreading of channel power in the delay-Doppler domain for each transmitter. %
		The delay is with respect to the \ac{LOS} component. %
		The ten largest peaks are shown as blue circles for the \acf{LSF}, and red circles for the \acf{SBL} results based on ten iterations. %
		The wireless channel is dominated by a single cluster in the delay-Doppler domain at the \ac{LOS} component. %
		The \ac{SBL} results agree with the largest peaks of the \acp{LSF}.
		Two strong paths with equal Doppler but different delay are resolved by \ac{SBL}, e.g.\ at \SI{2.6}{\second} recording time. %
	}
	\label{fig:delaydoppler}
\end{figure*}

The channel power spreading in delay-Doppler domain, over the course of the measurement, is shown in \cref{fig:delaydoppler}.
The spreading is mostly limited to a cluster around the \ac{LOS} component and is thus described as a sparse channel in the delay-Doppler domain.
Components arriving before the \ac{LOS} are due to the previously mentioned reflections beyond the crossing (aliasing), and \ac{DFT} leakage.
Results from the sparse channel estimation with \ac{SBL} are shown in \cref{fig:delaydoppler}. They are in good agreement with the peaks of the \ac{LSF}.
Furthermore, finer sampling in the delay domain reveals additional \acp{MPC}.

\section{Conclusion}
The measured \ac{V2I} wireless channels in an urban environment are dominated by one cluster in the delay-Doppler domain.
In a communication system, this allows for easy compensation for delay and Doppler shifts of the whole cluster, and enables the design of low-complexity receivers.
This type of channel is especially suitable for \ac{OTFS} modulation.
Both considered beam elevations of the antennas at the car's roof have only a weak impact on the \ac{V2I} channel in delay-Doppler domain.
However, the elevated beam is less subject to fading due to weaker illumination of the street level.
\ac{SBL} is able to find a sparse representation of the measured channel, which agrees with results from the \ac{LSF}, and reveals additional \acp{MPC}.

\clearpage
\bibliographystyle{IEEEtran}
\begin{acronym}
	\acro{AWG}{arbitrary waveform generator}
	\acro{SBL}{sparse Bayesian learning}
	\acro{DFT}{discrete Fourier transform}
	\acro{LSF}{local scattering function}
	\acro{LOS}{line of sight}
	\acro{DPSS}{discrete prolate spheroidal sequences}
	\acro{OTFS}{orthogonal time frequency space}
	\acro{DSD}{Doppler power spectral density}
	\acro{SA}{signal analyzer}
	\acro{MPC}{multipath component}
	\acro{mmWave}{millimeter wave}
	\acro{V2V}{vehicle-to-vehicle}
	\acro{V2I}{vehicle-to-infrastructure}
	\acro{SFFT}{symplectic finite Fourier transform}
	\acro{ISFFT}{inverse SFFT}
	\acro{TX}{transmitter}
	\acro{RX}{receiver}
	\acro{LO}{local oscillator}
	\acro{RF}{radio frequency}
	\acro{PLL}{phase-locked loop}
	\acro{CFO}{center frequency offset}
	\acro{SNR}{signal-power to noise ratio}
	\acro{ZC}{Zadoff-Chu}
	\acro{CAZAC}{constant amplitude zero autocorrelation}
\end{acronym}
\bibliography{references}
\end{document}